%% file: main-journal.tex
\documentclass[10pt,journal,compsoc]{IEEEtran}
\IEEEoverridecommandlockouts
\pdfoutput=1
\usepackage{cite}
\usepackage{amssymb}
\usepackage{mathtools}
\usepackage{amsmath}
\usepackage[utf8]{inputenc}
\usepackage[english]{babel}
\usepackage[algo2e]{algorithm2e}
\usepackage{graphicx}
\usepackage{epstopdf}
\usepackage{subfig}
\usepackage{textcomp}
\usepackage{xcolor}
\usepackage{booktabs}
\usepackage{enumitem}
\usepackage{tabularx}
\usepackage{algpseudocode}
\usepackage{algorithm}
\usepackage{multirow}
\usepackage{caption}
\usepackage{etoolbox}
\usepackage{multicol}
\usepackage{tabularx}
\usepackage{times, alltt, xspace}
\usepackage{setspace}
\def\BibTeX{{\rm B\kern-.05em{\sc i\kern-.025em b}\kern-.08em
    T\kern-.1667em\lower.7ex\hbox{E}\kern-.125emX}}

\makeatletter
\def\endthebibliography{%
  \def\@noitemerr{\@latex@warning{Empty `thebibliography' environment}}%
  \endlist
}
\makeatother

\newcommand{\effg}{\ensuremath{\mathrm{effS}}}
\newcommand{\effu}{\ensuremath{\mathrm{effS}}}

\newcommand{\Gr}{\ensuremath{\mathrm{G}}}
\newcommand{\vt}{{$\mathrm{V_T}$}}
\newcommand{\s}{{$\mathrm{s^\prime}$}}
\newcommand{\vp}{{$\mathrm{v^\prime}$}}
\newcommand{\ob}{{$\mathrm{tc^\prime}$}}

\newcommand{\evt}{EV$\mathrm{_T}$}
\newcommand{\range}{\ensuremath{\mathrm{Range(att)}}}
\newcommand{\dirug}{\ensuremath{\mathrm{directG}}\xspace}

\makeatletter
\long\def\@IEEEtitleabstractindextextbox#1{\parbox{0.922\textwidth}{#1}}
\makeatother
\begin{document}
\title{Secure V2V and V2I Communication in \\ Intelligent Transportation using Cloudlets}

\author{Maanak~Gupta, James Benson, Farhan Patwa
        and~Ravi~Sandhu,~\IEEEmembership{Fellow,~IEEE}
\IEEEcompsocitemizethanks{\IEEEcompsocthanksitem Maanak Gupta is Assistant Professor in the Department
of Computer Science, Tennessee Tech University, Cookeville,
TX, 38505, USA. \protect\\
E-mail:mgupta@tntech.edu}
\IEEEcompsocitemizethanks{\IEEEcompsocthanksitem James Benson, Farhan Patwa and Ravi Sandhu are with the Institute for Cyber Security and Department
of Computer Science, University of Texas at San Antonio, San Antonio,
TX, 78249, USA.
E-mail: james.benson@utsa.edu, farhan.patwa@utsa.edu, ravi.sandhu@utsa.edu}}

\markboth{IEEE Transactions on Services Computing, Vol xx, No xx, Month xx}%
{Shell \MakeLowercase{\textit{et al.}}: Bare Demo of IEEEtran.cls for Computer Society Journals}

\IEEEtitleabstractindextext{%
\begin{abstract}
Intelligent Transportation System (ITS) is a vision which offers safe, secure and smart travel experience to drivers. This futuristic plan aims to enable vehicles, roadside transportation infrastructures, pedestrian smart-phones and other devices to communicate with one another to provide safety and convenience services. Vehicle to Vehicle (V2V) and Vehicle to Infrastructure (V2I) communication in ITS offers ability to exchange speed, heading angle, position and other environment related conditions amongst vehicles and with surrounding smart infrastructures. In this intelligent setup, vehicles and users communicate and exchange data with random untrusted entities (like vehicles, smart traffic lights or pedestrians) whom they don't know or have met before. The concerns of location privacy and secure communication further deter the adoption of this smarter and safe transportation. In this paper, we present a secure and trusted V2V and V2I communication approach using edge infrastructures where instead of direct peer to peer communication, we introduce trusted cloudlets to authorize, check and verify the authenticity, integrity and ensure anonymity of messages exchanged in the system. Moving vehicles or road side infrastructure are dynamically connected to nearby cloudlets, where security policies can be implemented to sanitize or stop fake messages and prevent rogue vehicles to exchange messages with other vehicles. We also present a formal attribute-based model for V2V and V2I communication, called AB-ITS, along with proof of concept implementation of the proposed solution in AWS IoT platform. This cloudlet supported architecture complements direct V2V or V2I communication, and serves important use cases such as accident or ice-threat warning and other safety applications. Performance metrics of our proposed architecture are also discussed and compared with existing ITS technologies.\end{abstract}

}

\maketitle

\IEEEdisplaynontitleabstractindextext
\IEEEpeerreviewmaketitle
\input{Sections/Introduction}

\input{Sections/Related_work}
\input{Sections/Proposed_Arch}
\input{Sections/Model}

\input{Sections/Implementation}
\input{Sections/Summary}
\bibliographystyle{IEEEtran}
\bibliography{acmart}
\begin{IEEEbiographynophoto}{Maanak Gupta}
received his PhD in Computer Science from the University of Texas at San Antonio. He is currently assistant professor at Tennessee Tech University. His main areas of interest include security and privacy in cyberspace.
\end{IEEEbiographynophoto}
\vspace{-2 cm}
\begin{IEEEbiographynophoto}{James Benson}
is the Technology Research Analyst at the ICS at the University of Texas at San Antonio.
 He holds a masters in Electrical Engineering focusing on Computer Science from the UTSA and a masters in Physics from Clarkson University.
\end{IEEEbiographynophoto}
\vspace{-2 cm}
\begin{IEEEbiographynophoto}{Farhan Patwa}
is the Associate Director and Chief Architect at the ICS at the University of Texas at San Antonio.
\end{IEEEbiographynophoto}
\vspace{-2 cm}
\begin{IEEEbiographynophoto}{Ravi Sandhu}
is the founding Executive Director of the Institute for Cyber Security at the University of Texas at San Antonio, where he holds the Lutcher Brown Endowed Chair in Cyber Security.
He is a fellow of the ACM, IEEE and AAAS and an inventor on 31 patents.
\end{IEEEbiographynophoto}

\end{document}

%% file: Sections/Introduction.tex
\IEEEraisesectionheading{\section{Introduction and Motivation}\label{sec_intro}}
\IEEEPARstart{F}{uture} smart world will be equipped with technologies and autonomous devices which collaborate among themselves with minimal human interference. Automotive industry is one of the front runners that has quickly embraced this technological change. Connected vehicles (CVs) and smart cars have been introduced, with a plethora of on-board sensors and applications with internet connectivity to offer safety and comfort services to users. Intelligent transportation for smart cities envision moving entities interacting and exchanging information with other vehicles, infrastructures or on-road pedestrians. Federal and private agencies are defining communication standards and technologies for Intelligent Transportation System (ITS) to ensure safety, and address security and privacy concerns of end users. 

Vehicle to Vehicle (V2V) and Vehicle to Infrastructure (V2I) are two proposed technological innovations which can change current transportation. V2V will enable vehicles to exchange information about speed, location, position, direction, or brake status with other surrounding vehicles where receiving vehicles will aggregate these messages and make smart decisions using on-board applications which will warn drivers about accidents, over-speed, slow traffic ahead, aggressive driver, blind spot or a road hazard. V2I will enable road side units (RSUs) or traffic infrastructures to transmit information about bridge permissible height, merging traffic, work zone warning or road hazard detection to complement V2V applications. Vehicle to pedestrian (V2P) is also envisioned to cater to pedestrians, such as with visual or physical impairments, and send corresponding alerts to approaching vehicles. These communication technologies will use Dedicated Short Range Communications (DSRC) \cite{dsrc} to exchange data packets, called Basic Safety Messages (BSMs) \cite{bsm}, with nearby vehicles and entities between 300-500 meters range. Messages will be sent up to 10 times per second providing a 360-degree view of proximity, with on-board applications using the information for triggering alerts and warnings. US Department of Transportation (DOT) and National Highway Traffic Safety Administration (NHTSA) estimate around 80\% of non-impaired collisions \cite{us-dot,us-dot-1} and 6.9 billions traffic hours can be reduced by using V2V, V2I and V2P communications.

Vehicles in ITS are communicating and exchanging information with external entities including toll booths, gas stations, parking lots, and other vehicles, which raises security and privacy issues. Incidents on Jeep and Tesla \cite{jeep,tesla} have been demonstrated where car engine was shut and steering wheel controlled remotely by adversaries. These smart cars are equipped with 100's of electronic control units (ECUs) and more than 100 million lines of code, thereby, exposing broad attack surface for critical car systems including transmission control, air-bag, telematics, engine or infotainment systems. In-vehicle controller area network (CAN) bus also needs security to prevent unauthorized data exchange and tampering among ECUs and software manipulation. Cyber attacks on smart connected vehicles \cite{gao,nhtsa-1,nhtsa-2,elmaghraby2014cyber} include: unauthorized over the air updates (OTA) for firmware, stealing user private data, spoofing sensors, coordinated attacks on road side infrastructure or malware injection. Dynamic and mobile nature of V2X (Vehicle to everything) communication makes it additionally difficult to secure the distributed system where vehicles will be exchanging data with random unknown entities on road. Impersonation and fake message from malicious compromised vehicles is also a grave concern as the information exchanged is used by other vehicles to make alerts and notifications. Vehicle users also have privacy concerns where every movement can be tracked continuously by agencies or data collected from vehicles can be used to extrapolate personal identifiable information (PII).  These concerns lead to reluctance in embracing these future transportation technologies.

Attribute-based access control (ABAC) \cite{jin2012unified,hu2013guide,gupta2016mathrm} provides fine grained authorization capabilities for resources in a system. This mechanism offers flexibility in a distributed multi-entity dynamic environment where the attributes of entities along with contextual information are used to make access and communication authorization decisions. Intelligent transportation system involves interaction and messages exchange among entities with no prior association. Attributes of vehicles or transportation infrastructure can be used to authorize communication decision based on their current location, ownership or degree of trust. Such security mechanisms can help to prevent fake messages, stop rogue vehicles and ensure privacy aware message communication besides ensuring location and time sensitive relevance of exchanged information.

In this work, we present a privacy-aware secure attribute-based V2V and V2I communication architecture and model using trusted cloudlets. These cloudlets are setup in wide geographic locations with defined coverage area.  Each cloudlet will receive messages from vehicles in its range and forward it to all other vehicles associated with that cloudlet. Vehicles are dynamically assigned to these cloudlets as they move along geographic boundaries based on their GPS coordinates and predicted path. An important benefit of this indirect V2V and V2I communication is the deployment of security policies at edge cloudlets which can restrict or block fake messages, and ensure trustworthiness in communication. Moreover edge cloudlets also enable message anonymization and user privacy, as the receiver cannot detect who is the sender as all messages come through edge infrastructures. These cloudlets can also be used to forward certificate revocation lists (CRLs) to vehicles in the range beside blocking the vehicles themselves. Rogue vehicle list can be dynamically updated at the edges, and messages from a vehicle in the rogue list can be blocked. The proposed architecture and attribute-based policies ensure the important security properties of message integrity, originator authenticity and user privacy concerns in V2V and V2I communication. This MQTT \cite{mqtt} based approach for messages exchange can be used in addition to DSRC to enable use cases with acceptable latency (discussed in implementation section) without the need for additional hardware cost\footnote{NHTSA proposed V2V equipment and communication is between \$341 to \$350 per vehicle in 2020 \cite{faq}} and work with familiar technologies such as WiFi, LTE or 5G. This work proposes a formalized communication security model for V2V and V2I called attribute-based intelligent transportation system (AB-ITS). We have implemented our proposed architecture and model using AWS and collected several performance metrics, which reflect the plausibility and efficiency of our proposal.

Rest of the paper is as follows: Section \ref{sec_related} discusses related work along with USDOT proposed Security Credential Management System (SCMS). Security and privacy requirements along with the proposed cloudlet supported ITS architecture is given in Section \ref{sec_arch}. Section \ref{sec_model} presents formal attribute-based V2V and V2I communication model (AB-ITS). Section \ref{sec_imp} describes our implementation with real-world use cases using AWS, and discusses performance parameters. Section \ref{sec_summary} concludes the paper.

%% file: Sections/Related_work.tex
\section{Related Work}\label{sec_related}
Connected and smart vehicle applications need wireless exchange of V2X messages among unknown moving vehicles, RSUs and pedestrians. The proposed intelligent transportation system (ITS) for future cities has underlying technologies, security concerns and proposed solutions, which we briefly review in this section.

\subsection{Security Credentials Management System}
United States Department of Transportation (USDOT) has suggested a PKI-based security infrastructure system, called Security Credentials Management System or SCMS \cite{its-scms2, its-scms1}, to ensure trusted V2V and V2I communication among random moving entities. Authorized participating vehicles use digital certificates issued by SCMS to validate and authenticate basic safety messages (BSMs), by attaching these certificates with each message to ensure integrity, confidentiality and privacy of the communication. Vehicles need initial enrollment into SCMS to obtain security certificates from trusted certificate authorities (CA). Each BSM will include vehicle related information digitally signed using private key corresponding to the digital certificate attached with BSM. Different certificate types are used including enrollment, pseudonym and identification for vehicle and enrollment applications for RSUs. Certificates can be cancelled for potential adversaries or reported misbehaving vehicles by CAs by disseminating certificate revocation lists (CRLs). USDOT and NHTSA claim \cite{us-dot-1} that BSMs will exchange anonymized information and no personal identifiable data will be shared with other entities. SCMS is considered as a central system to be trusted by entities participating to revolutionize transportation.

However, there are some challenges \cite{scms-issues,scms-issues1} that need to be addressed before the system is deployed. Each vehicle will receive 20 certificates weekly to sign the BSMs \cite{nhtsa-5}, which will rotate every 5 minutes. Therefore, a vehicle will use a new set of 20 certificates every 100 minutes. In such a scenario a computer can analyse all the certificates a vehicle used in a day and then use these certificates to track it for a week. Although, PKI based SCMS system ensures who signed the certificate, it is difficult to prove how correct or true the information sent  from the vehicle is. A malfunctioning device in the vehicle can result in false BSMs exchanged even though the sender is trusted. Further, the proposed SCMS system will be largest and complex ever built producing 265B to 800B certs/year depending on weekly rate supporting 17M vehicles/year \cite{scms-issues1}. The revocation of certificates for bad actors would result in pushing CRLs to all enrolled vehicles, which will be time and bandwidth consuming.

\subsection{Relevant Background and Technologies}

Several general IoT architectures \cite{atzori2010internet,gubbi2013internet,alshehri2016access} have been proposed with different middleware layers in multi-layer stack representing physical objects, communication or service layer, cloud and end-user applications. Gupta and Sandhu proposed \cite{gupta2018authorization} enhanced access control oriented architecture (E-ACO) particularly relevant to smart cars and intelligent transportation. The work introduced clustered objects (smart objects with multiple sensors like cars) as component of object layer which interact with other objects similar to V2V and V2I communication. As shown in Figure \ref{fig_eaco}, E-ACO architecture has four layers: \textbf{Object Layer} at the bottom representing physical objects including connected cars, vehicles and RSUs. \textbf{Virtual Object Layer} maintains cyber entity (like an AWS shadow stored as JSON) of each physical object which is imperative in a moving and dynamic ecosystem like smart cars, where the connectivity of a vehicle is not continuously guaranteed. With virtual objects, when direct communication with physical object is not possible, its virtual entity maintains last reported and desired state information. Further, it resolves the issues of heterogeneity as objects support different communication technologies. Using virtual objects, physical entities communicate with corresponding virtual objects where messages are exchanged with virtual entities of other object which is then passed to actual physical object. \textbf{Cloud Services and Application Layer} together harness data sent by physical objects and use it to extrapolate value, analytics and provide end user cloud supported applications.

\input{Figures/v2v}

Smart cars security incidents including Jeep \cite{jeep} and Tesla Model X \cite{tesla} hacks have demonstrated how engine was stopped
and steering remotely controlled exhibiting cyber threats. Security and privacy issues in smart cars and ITS are serious concerns where several federal agencies are working along with industry partners to ``fully'' proof the system before final deployment and use by common public. European Union Agency for Network and Information Security (ENISA) \cite{enisa} has studied vulnerable assets in smart cars with related threat and risks, and proposed some prevention approaches with recommendations. Cooperative Intelligent Transport Systems (C-ITS) \cite{c-its1,c-its2} also highlighted the need of data communication integrity and authenticity in V2V and V2I, and proposed PKI based trust model using pseudonym certificates. NHTSA report \cite{nhtsa-3} has thoroughly explored the technical, legal and policy related issues pertinent to V2V communication and studied technological solutions for safety and privacy issues. US Government Accountability Office (GAO) \cite{gao} has also discussed security risks and potential attack surfaces in smart vehicles, and  proposed solutions to prevent cyber threats.

Attribute based access control \cite{hu2015attribute,jin2012unified,gupta2018attribute} provides fine grained authorization capabilities most appropriate in dynamic and distributed systems similar to ITS. Recently dynamic groups and ABAC model \cite{Gupta:2019:DGA:3292006.3300048,gupta2019secure} was proposed for smart cars ecosystem which caters to mobile needs of vehicles. However the model is more suitable to cloud assisted applications and a real time V2V and V2I edge supported model is still missing. Role based access controls \cite{sandhu1996role,ferraiolo2001proposed} were designed particularly for enterprise applications with a limited set of roles and administrators assigning roles to users. Similar concept does not seem to fit dynamic and random unknown IoT smart cars setting where devices and vehicles are in different administrative domains spread across geographic area.


%% file: Figures/v2v.tex
\begin{figure}[t]
\centering
\includegraphics[scale = .50]{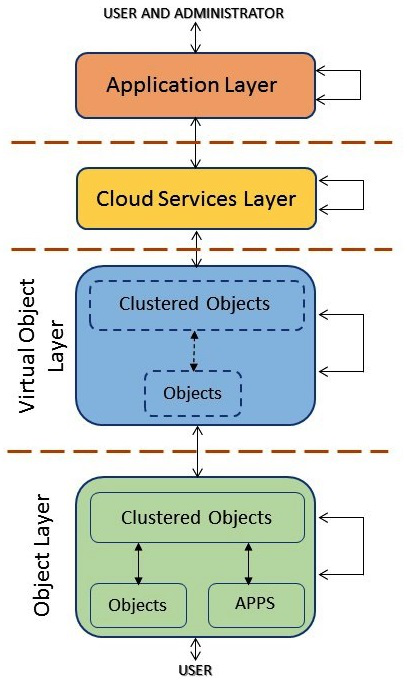}
\caption{Four Layered E-ACO Architecture \cite{gupta2018authorization}}
\label{fig_eaco}
\end{figure}
\begin{figure}[t]
\centering
\includegraphics[scale = .65]{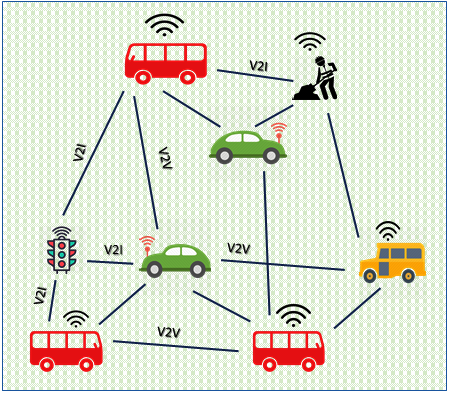}
\caption{V2V and V2I Peer to Peer Communication}
\label{fig_v2v}
\end{figure} 

%% file: Sections/Proposed_Arch.tex
\section{Proposed Cloudlets Supported ITS Architecture}\label{sec_arch}
The current peer to peer V2V and V2I communication as represented in Figure \ref{fig_v2v} is proposed to use SCMS to ensure secure trusted basic safety messages exchange among entities. However, the vast and complex scale of this PKI based system has user privacy and security concerns which need to be addressed before its deployment. In this section, we will discuss security and privacy requirements of ITS and smart cars ecosystem and highlight how the proposed trusted cloudlets supported communication offers the required security and complements current solutions.

\subsection{Security and Privacy Requirements}
\input{Figures/architecture}
Intelligent Transportation System (ITS) involves real time sharing of location and sensitive information about vehicles and passengers, which pose a serious privacy threat and a strong deterrent for its adoption. Dynamic and distributed ITS will enable interaction with random entities on road with no prior trust established, and the information sent from these smart vehicles will be used by on-board applications to provide safety and warning signals, which itself has some inherent security risks. An adversary can compromise a road-side unit or vehicle to send fake information about traffic or accident, which can trigger unnecessary alerts and may distract drivers. Basic safety messages (BSMs) are designed to contain no personal identifiable information (PII) and are attached with a certificate issued by certificate authority in SCMS. However, limited number of certificates and number of messages sent per minute can reveal the identity of a targeted vehicle with advanced computer techniques. Untrackability of vehicles and users is paramount to ensure privacy in ITS. Also, the system must not save personal or individual information and use it as law enforcement or issuing speeding tickets. Anonymity of sender must always be maintained. Over the air messages exchanged among smart entities must have integrity, and authenticity. Security mechanisms to protect smart cars and their critical systems from unauthorized access, control and tampering are important to strengthen intelligent transportation. Integrated approach of DSRC and cellular technologies is needed based on different ITS applications. Cloud and cloudlets supported architectures will provide resiliency and reduce system stress. Encrypted and secure data transfer link is the backbone needed from DSRC, cellular LTE or any communication technologies involved in ITS. However, limited bandwidth and latency issues in cloud connectivity needed for certificate updates and revocation needs attention.

In smart city, location based notifications for connected vehicles must allow user to have personal preferences where a user may want weather warning and not parking advertisements on board. Dynamic policies are required, for example, in case of a traffic jam in an area a policy may ask all drivers to follow route A but considering the heavy traffic on route A, the policy may be changed to move traffic to route B or C. This can be implemented at the edge level and triggered by central administrators. In such a case, whether the administrative subject is authorized to change the policy or trigger an alert, also needs security checks.

\subsection{How Cloudlets Can Provide Security?}
Figure \ref{fig_arch} shows the proposed edge supported architecture for V2V and V2I communication. Trusted edge infrastructures (setup by city administration) will work as a middle man and relay messages to vehicles and other entities inside its geographic range. Instead of peer to peer connection, all vehicles publish to edges, where security policies defined are checked to ensure validity and integrity of the communication, and relevance of messages, before forwarding to other vehicles. A vehicle can be in range of multiple infrastructures, depending on its location.  Each vehicle will be dynamically associated with edges as it moves. All participating vehicles and RSUs still need to enroll with a central authority to be part of the system, to ensure that only trusted vehicles are allowed to exchange messages among themselves. Communication technologies used for vehicles to cloudlets can be cellular LTE, WiFi or DSRC. MQTT messaging protocol can be used, as discussed in implementation which will obviate the cost of DSRC equipments needed in smart cars. The proposed architecture is implemented in addition to V2V and V2I direct communication and is supported in NPRM \cite{nhtsa-6} documents which recommend both DSRC and secondary communication for ITS.

Trusted cloudlets installed in wide geographic area offer the needed fog infrastructure functionality required in an IoT environment. They can address security concerns by deploying and enforcing security policies to ensure trusted communication among smart entities on the road. 
This proposed architecture offers an alternate edge supported V2V and V2I communication with minimal message latency and in permissible time limits \cite{Xu:2004:VSM:1023875.1023879,articleV2V}.
A vehicle sending and receiving BSM or other messages, must be associated with an edge infrastructure, which will enforce policies, sanitize messages, prevent fake messages dissemination and offer administrative advantages. Each cloudlet will have a geographic range and all the vehicles within it will get associated with the edge automatically. Since the range of edge is within a restricted limited area, it also ensures location sensitivity of messages exchanged, as vehicles communicating messages must be associated to a common edge cloudlet. Message anonymity and sanitization can be done, since
the messages sent by a vehicle are relayed via the edge cloudlet without direct peer to peer communication, which will have less security and privacy implications.

Further, using cloudlets offers administrative benefits as single notification from edge infrastructure will trigger alerts for all the vehicles which are connected to it in a geographic range. If an agency or a police vehicle wants to send alerts, instead of sending to each individual vehicle, they can send it to a trusted cloudlet, which after checking the policies to ensure the sender is allowed to generate such requests, forwards or stops the message. 
Also, entities present in a particular area have certain characteristics (for example, stop sign warning, speed limits, deer-threat, flash flood warnings etc.) in common, which can be inherited by getting dynamically associated to edge infrastructures, without the need to generate messages 10 times per second \cite{nhtsa-3} to get this information from other vehicles or RSUs saving network bandwidth.

It is also possible to limit the messages to a specific set of vehicles, for example, in case of a kidnapped child warning, messages can be sent to nearby edge infrastructures and then to only police vehicles in the area, and not to the common public using security policies defined at the cloudlet. Edge infrastructure can also have the capacity to filter unwanted and incorrect messages from the vehicles and infrastructure using a majority rule policy.  For example, if an adversary is sending accident message (either deliberately or a malfunction sensor on vehicle) to subvert the traffic whereas other vehicles notify no accident and clear traffic messages, installed trusted edge will have the intelligence and policy to filter such fake messages and forward the correct information to its associated vehicles.
This will not be possible in peer to peer V2X (vehicle to anything) architecture immediately, until certificate revocations (by a central authority) are propagated to individual vehicle, which may take time and also require internet connectivity which cannot be guaranteed all times in terrains where the vehicle is moving. Also, instead of sending CRLs to each vehicle, only edge servers can be sent with list of revoked certificates and based on the information, edge can decide if the messages sent by vehicle should be forwarded or not.

Further, if an adversary is detected by an edge with fake or wrong messages, policies can be defined to inform appropriate agencies and law enforcement in the area where such malicious behaviour is detected. It is also possible to have different levels of alerts based on the degree of trust and who is the sender. Law enforcement initiating a bomb threat in the vicinity will be treated as major threat and edge infrastructure states it as code red alert, with immediate rerouting and emergency exit directions.

%% file: Figures/architecture.tex
\begin{figure*}[t]
\centering
\includegraphics[scale = .50]{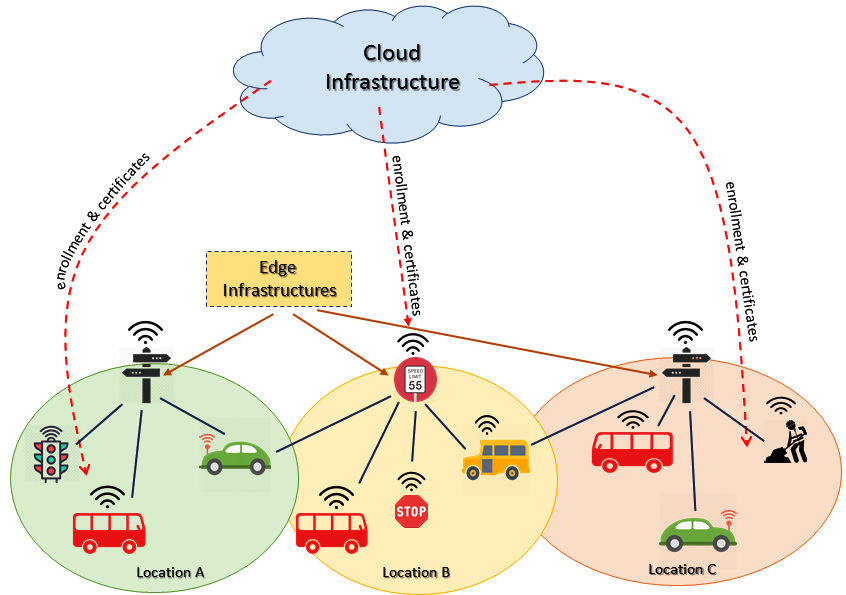}
\caption{Proposed Trusted Cloudlets Supported V2V and V2I Communication Architecture}
\label{fig_arch}
\end{figure*} 

%% file: Sections/Model.tex
\section{ Cloudlets Enabled Attribute Based V2V and V2I Communication}\label{sec_model}
Edge cloudlets supported V2V and V2I communication has many advantages, as discussed in previous section. These cloudlets can support attributes based fine-grained policies based on which communication decisions can be made. These attributes offer flexibility and take into account different environmental factors along with dynamic policies based on administrator needs. Further, individual users are also allowed to set their own privacy preferences, to decide on what and from whom messages are allowed to receive. In this section, we formally define our proposed cloudlets supported attributes based intelligent transportation system model, which we refer to as AB-ITS.

\subsection{AB-ITS Communication Model}
\input{Figures/model}
The conceptual AB-ITS communication model is shown in Figure \ref{fig_model} and formal definitions elaborated in Table \ref{tab 1}. The model has following components: Vehicles (V), Transportation Infrastructure Devices (I), Users (U), Sources (S), Trusted Cloudlets (TC), Target Vehicles (\vt), Operations (OP), Authorization Policies (POL), and Attributes (ATT).

\noindent
\textbf{Sources (S) :} A source initiates operations on cloudlets (discussed below) in the system. A source can be from a set of vehicles (V), transportation infrastructure (I) or an administrator user (U). For instance, in case of V2V communication, a source is a vehicle which wants to send messages to other vehicles in its vicinity. Similarly, law enforcement and city administration can initiate theft and accident alerts in a particular area via cloudlets, which are forwarded to all vehicles associated with cloudlet.

\noindent
\textbf{Trusted Cloudlets (TC) :} Cloudlets are introduced, which are trusted edge infrastructures set up across locations and facilitate secure V2V and V2I communication. These cloudlets have a limited geographic range and all vehicles in that range get associated with one or more TCs automatically based on their moving location coordinates. Any communication between vehicles and other entities including transportation infrastructures (or RSUs) is done via TC, which checks security policies to forward or block the messages sent by different sources. Also TCs have attributes which are propagated to associated vehicles and can also help setting alerts and warnings based on attribute values. For instance, when a vehicle enters forest and gets associated with the cloudlet, it can automatically inherit a wildlife area attribute ON from TC.

\noindent
\textbf{Target Vehicles ($\mathbf{V_T}$) :} These vehicles are subset of total vehicles (V) in the transportation system and are potential receiver of messages sent by a source. Both target vehicles and source must be associated with same TC to enable V2V and V2I communication.

\noindent
\textbf{Operations (OP) :} Operations are actions which are performed by source on TC. TC also execute operations against associated vehicles and infrastructures. For example, a source initiating a join operation to get associated with a TC, or trying to send a message to vehicles via TC. Also, TC forwarding a message sent by sources to its member vehicles is another example of operations in ITS. These also include administrative actions performed by a user including updating, deleting or adding attribute values for an attribute or rogue vehicles list in TC.

\noindent
\textbf{Authorization Policies (POL) and Attributes (ATT) :} Sources, TCs, vehicles and other relevant ITS entities can have personal defined individual policies along with system wide authorization policies needed for the overall secure functioning of the ecosystem. Vehicles owners can set individual privacy preferences which enable them to allow or disallow any particular private information from being shared with a third party remotely. Similarly, city traffic department may set its own rules when to trigger an alert or warnings to vehicles in a sensitive or accident prone area. Administrative policies are also needed to authorize a legitimate user to change attributes, send notifications to TCs or update rogue vehicles list. Entities like vehicles and sources also have individual characteristics, called attributes, which are used to make authorization and communication decisions in ITS. For a vehicle, sample attributes can be: vehicle ID, speed, heading angle, brake, vehicle size, vehicle type or preferred notifications. Vehicles and infrastructure can also inherit attributes from their associated TCs, which can have common location wide attributes like speed limit, road work ahead or blind turn.

Both attributes and policies are dynamic which can be changed by administrators or vehicle owners based on system needs and personal preferences. The attributes of vehicles like location, speed or heading angle are continuously changing, but other attributes like vehicle size remain static. Policies are also dynamic in nature, as reflected in use-case implementation in the next section, where we defined a security policy with a list of black-listed rogue vehicles which are notified to law enforcement when detected by TCs. This list is dynamic in nature and is continuously updated by administrators, demonstrating how dynamic policies are used and enforced in ITS.
It must be noted that in a session the proposed model assumes a static set of policies and attributes which are used to make V2V and V2I communication decision. All relevant polices including system defined and user preferences are evaluated to make the final communication decision.
\input{Tables/table_model-new}

In our proposed model, TCs evaluate security policies and ensure that un-trusted or fake messages are not forwarded to associated vehicles in its geographic coverage boundary. These connected vehicles must initiate association with TCs pro-actively based on their predicted path, and once they get into the range of the TCs, vehicles become the member of TC. Such communication with TCs can be done using encrypted and secure cellular or WiFi technologies with no added equipment cost. It should be noted that our model complements the proposed DSRC based direct V2V and V2I communication, and can be used to assist in situations where the authenticity and integrity of messages is much needed. Our use-cases in the next section will highlight the real world enforcement and use of AB-ITS model.

\subsection{Formal Definitions}

Table \ref{tab 1} elaborates the formal AB-ITS communication model definitions, which comprise of vehicles (V), transportation infrastructure devices (I), administrative users (U) and edge cloudlets (TC). A source in S initiating an operation op $\in$ OP can be from a set of vehicles, transportation infrastructures or users, whereas target vehicles \vt~is a subset of total vehicles in the entire transportation system (\vt~$\subseteq$ V). Attributes are functions defined for source and edge cloudlets, where functions can be set or atomic valued (stated by attType) and are assigned values from Range(att) for each att $\in$ ATT. The atomic valued attributes are assigned single value including null (denoted as $\bot$) whereas set valued attribute can have a subset of values assigned from power set of the range of attribute function. Some attributes are also defined system wide, which reflect the state of entire transportation system (like level of threat or city traffic) and are set by administrators. Authorization security policies are defined for individual sources and TCs, which are either stated based on personal privacy preferences or are enforced system wide as defined by central administrators. For example, a driver may not want to receive marketing commercials on dashboard, so she can set such personal preference as choosing the desired policy, whereas police can define a policy with a list of black-listed cars and blocking communication from them.

Source and target vehicles are dynamically assigned to one or many trusted edge cloudlets based on their current GPS coordinates and predicted path as defined by associated\_cloudlets function. The association with edge cloudlets is fixed for transportation infrastructures or administrators which are assigned at the time of system deployment whereas for vehicles it keeps on changing as the vehicles move. Each cloudlet has defined geographic coverage area and when vehicles enter the area, they get associated with the cloudlet. A vehicle may be associated with multiple cloudlets in areas where coverage areas are overlapping, thereby, a vehicle is always associated to at least one cloudlet at all times. These cloudlets mediate the V2V and V2I communication by enforcing security policies, stop fake messages and ensure privacy, as discussed later in the model definitions. Further, sources (including vehicles) inherit attributes from their associated cloudlets, which helps in administration and propagation of common attributes to all associated entities with single administrative action. For instance, at a location where flash flood warning is issued, the edge cloudlet installed there will set attribute flash-flood = ON for all its associated vehicles when they become members of that cloudlet. In case of set valued attribute function, the effective attribute values for att $\in$ ATT of source (defined as $\mathrm{\effu_{att}}$), including target vehicles, is the union of direct values assigned to the source for attribute att and the values assigned to att for each associated cloudlets. However, in case of atomic valued attribute, it is necessary to define which attribute values take precedence when multiple edge clouds are associated. In our model, we propose that most recently connected cloudlet with non null value for the attribute will be inherited by the associated source or vehicles.\footnote{There are other approaches also to deal with atomic value inheritance, but for moving vehicles which are dynamically assigned to new cloudlets, we believe this approach is the most appropriate and relevant.} For example, the speed-limit attribute of most recently associated cloudlet will be populated for all member vehicles, and as the vehicle moves, this value is inherited from next associated edge cloudlet and so on. This inheritance in atomic values attribute only takes place when edge cloudlets have non null values, whereby with all associated cloudlets having null values, the direct attribute value of the source holds as its effective value also.

Authorization functions are parameterized propositional logic formulae defined to represent access control security policies stated in the policy language defined in Table \ref{tab 1}. The function $\mathrm{Auth_{op}}$(s:S, tc:TC) specify conditions under which source s (including vehicles) can perform an operation op $\in$ via cloudlet tc $\in$ TC. These boolean authorization functions are evaluated substituting actual arguments for formal parameters along with direct and effective attributes values of actual arguments. Similar syntax and policy language can be defined for other set of policies including personal vehicle specific policies or system wide policies with attributes of relevant entities substituted in authorization requests evaluation.
Authorization decision to allow \s~$\in$ S to perform an operation op $\in$ OP on \ob~$\in$ TC is determined when the authorization function is evaluated with the actual arguments (\s~$\in$ S, \ob~$\in$ TC) to be True.
Similarly, the decision for operation op from \ob~$\in$ TC to \vp $\in$ \vt~is made by calling the relevant authorization function with actual parameters.

As discussed in authorization property, the model has defined two primitive operations, `send' and `forward' relevant for V2V and V2I communication. A source uses `send' operation (defined as  $\mathrm{Auth_{send}}$(\s : S, \ob : TC)) to communicate a `send message' to trusted cloudlet, whereas `forward' operation (defined as $\mathrm{Auth_{forward}}$(\ob : TC, \vp : \vt)) is between trusted cloudlet and target vehicle defining a `forward message'. For allowing, communication from \s to \vp requires a common \ob~to which both \s and \vp are associated and the required authorization functions for send and forward messages i.e  $\mathrm{Auth_{send}}$(\s : S, \ob : TC) and $\mathrm{Auth_{forward}}$(\ob : TC, \vp : \vt) as well as the system defined security policies evaluate to True. Additional relevant operations and messages can be similarly defined.

The proposed AB-ITS model leverages attributes and GPS coordinates of communicating entities to enable and secure V2V and V2I communication. The introduction of trusted cloudlets provide benefits of enforcing security policies at the edge to stop fake messages, enhance user privacy and integrity of messages before forwarded to other target vehicles. These edge cloudlets ensure low latency and near real time communication much needed in most ITS applications without bandwidth issues. It must be noted that the messages shared among source and vehicles are end to end encrypted and can still use the proposed DSRC wireless technology for communication with cloudlet and then to the vehicles. Our model complements the USDOT proposed V2V and V2I architecture functionalities and support applications which need additional message integrity and confidentiality, and can be used as an add on to current ITS peer to peer communication. 

%% file: Figures/model.tex
\begin{figure}[t]
\centering
\includegraphics[scale = .70]{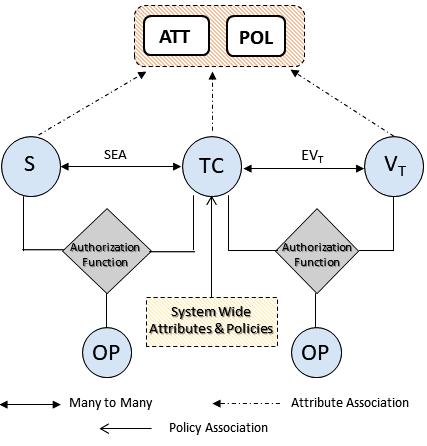}
\caption{A Conceptual AB-ITS Communication Model}
\label{fig_model}
\end{figure} 

%% file: Tables/table_model-new.tex
\begin{table*}[ht!]
\centering
\caption{Formal AB-ITS Communication Model Definitions}
\label{tab 1}
\renewcommand{\arraystretch}{1.1}
{%
\begin{tabularx}{\textwidth}{@{}llllll@{}}
\toprule
\multicolumn{6}{l}{\textbf{Basic Sets and  Functions}}\\
\multicolumn{6}{l}{-- V, I, U, TC are finite sets of vehicles, transportation infrastructure devices, (administrative) users, and trusted cloudlets respectively.}                              \\
\multicolumn{6}{l}{-- S, \vt, OP are finite sets of sources, target vehicles, and operations respectively, where S = I $\cup$ V $\cup$ U and \vt~$\subseteq$ V.}\\
\multicolumn{6}{l}{-- ATT is a finite set of attributes associated with S, TC, and system-wide.}                                                      \\
\multicolumn{6}{l}{-- For each attribute att in ATT,  Range(att) is a finite set of atomic values.}
\\
\multicolumn{6}{l}{-- attType: ATT = \{set, atomic\}, defines attributes to be set or atomic valued.}                                                      \\
\multicolumn{6}{l}{\begin{tabular}{@{}l@{}}-- Each attribute att in ATT maps entities in S and TC, and system-wide~to attribute values. Formally,\\\;\; att : S $\cup$ TC $\cup$ \{system-wide\} $\rightarrow$\; $ \begin{cases} \mathrm{{Range(att)}}~\cup \{\bot\} & \text{if attType(att) = atomic}\\ \mathrm{2^{Range(att)}} & \text{if attType(att) = set} \\ \end{cases}$ \end{tabular}}
\\
\multicolumn{6}{l}{-- POL is a finite set of authorization policies associated with individual entities in S and TC, and system-wide.}                                                      \\
\multicolumn{6}{l}{\begin{tabular}{@{}l@{}}-- associated\_cloudlets : $ \mathrm{S} \rightarrow$ 2$^\mathrm{TC}$,  maps each source (including target vehicles) to a set of trusted cloudlets.\\\;\; Equivalently, relations SEA = $\{(\mathrm{s,tc})\mid \mathrm{tc}\in \mathrm{associated\_cloudlets(s)}\}$ and
 \evt~ = SEA $\cap$ \vt~$\times$ TC.
 \end{tabular}}
\\
\multicolumn{1}{l}{\textbf{Effective Attributes of Sources Including Vehicles}}                                                                 \\
\multicolumn{6}{l}{-- For each attribute $\mathrm{att}$ in ATT such that attType(att) = set :}\\
\multicolumn{6}{l}{\begin{tabularx}{\textwidth}[c]{@{}l@{}}$ \;\;\;\; \bullet \;\;$ $\mathrm{\effu_{att}}$ : S~$\rightarrow {2^{\range}}$, defined as $\mathrm{\effu_{att}(s)}$ = $\mathrm{att(s)}$
$\bigcup_{\mathrm{tc\in associated\_cloudlets(s)}}$
$ \mathrm{att(tc)}$. \end{tabularx}} \\
\multicolumn{6}{l}{\begin{tabular}[c]{@{}l@{}}-- For each attribute $\mathrm{att}$ in ATT such that attType(att) = atomic : \\ $ \;\;\;\; \bullet \;\;$ $\mathrm{\effg_{att}}$ : S~$\rightarrow {{\range}} \cup \{\bot\}$, defined as \\\;\;\;\;\;\;\;\; $\mathrm{\effg_{att}(s)}$ = $ \begin{cases} \mathrm{att(s)} & \text{if $\forall~\mathrm{tc} \in \mathrm{associated\_cloudlets(s)}.~ \mathrm{att(tc)} = \bot, \mathrm{otherwise}$} \\ \mathrm{att(tc)} &
\text{where $\mathrm{tc}$ was most recently assigned $\mathrm{att(tc)} \neq \bot$ amongst all $\mathrm{tc'} \in \mathrm{associated\_cloudlets(s)}$}
\end{cases}$ \end{tabular}} \\

\midrule

\multicolumn{2}{l}{ \textbf{Authorization Functions (Policies)}}                                                                                                          \\
\multicolumn{6}{l}{\begin{tabularx}{\textwidth}[c]{@{}l@{}}-- {Authorization Function:} For each op $\in$ OP, $\mathrm{Auth_{op}(s:S, tc:TC)}$
is a parameterized propositional logic  formulae returning true\\\;\; or false, defined  using the following policy language: \end{tabularx}}\\
\multicolumn{6}{l}{\begin{tabular}[c]{@{}l@{}}$\;\;\;\; \bullet \;\; \mathrm{ \alpha \Coloneqq \alpha \land \alpha \;|\; \alpha \vee \alpha \;|\; (\alpha) \;|\; \neg \alpha \;|\; \exists\; x \in set.\alpha  \;|\; \forall\; x \in set.\alpha \;|\;  set \bigtriangleup set \;|\; } $ $ \mathrm{atomic \in set \;|\;  atomic \notin set }$\end{tabular}} \\

\multicolumn{6}{l}{$ \;\;\;\; \bullet \;\; \bigtriangleup \Coloneqq  \; \subset \;|\; \subseteq \;|\; \nsubseteq \;|\; \cap \;|\; \cup $}\\

\multicolumn{6}{l}{$ \;\;\;\; \bullet \;\;\mathrm{  set \Coloneqq {eff}_{att}(i) \;|\; {att}(i) }$}for $\mathrm{att}$ $\in $ ATT, i $\in$ S $\cup$ TC $\cup$ \{system-wide\}, attType(att) = set \;\;\;\;\;\;\;\;\;\;\;\;\;\;\;\;\;\;\;\;\;\;\;\;\;\\
\multicolumn{6}{l}{$ \;\;\;\; \bullet \;\; \mathrm{ atomic \Coloneqq {eff}_{att}(i) \;|\; {att}(i) \;|\; value }$} for $\mathrm{att}$ $\in $ ATT, i $\in$ S $\cup$ TC $\cup$ \{system-wide\}, attType(att) = atomic \;\;\;\;\;\;\;\;\;\;\;\;\;\;\;\;\;\;\;\\
\multicolumn{6}{l}{\begin{tabular}[c]{@{}l@{}}-- {Authorization Function Evaluation:} Authorization functions are evaluated by substituting actual arguments for formal parameters \\\;\; along with attribute values of actual arguments, thus reducing the parameterized formula to a propositional logic formula \\\;\; for evaluation. \end{tabular}}\\

\midrule
\multicolumn{2}{l}{ \textbf{Authorization Decision}}                                                                                                          \\
\multicolumn{6}{l}{\begin{tabular}[c]{@{}l@{}}-- A source \s $\in$ S is allowed to perform an operation op $\in$ OP on edge cloudlet \ob $\in$ TC (where \s and \ob are actual arguments),\\\;\; if all the required policies stated in $\mathrm{Auth_{op}}$(\s : S, \ob : TC), are satisfied. Formally, $\mathrm{Auth_{op}}$(\s : S, \ob : TC) = True.\end{tabular}}\\

\midrule
\multicolumn{2}{l}{ \textbf{Authorization Communication Property }}                                                                                                          \\
\multicolumn{6}{l}{\begin{tabular}[c]{@{}l@{}}-- To communicate a message between \s $\in$ S and  \vp $\in$ \vt~ requires authorization for individual operations \{send, forward\} $\in$ OP\\\;\; where  send[\s, \ob] and forward[\ob, \vp] are operation signatures meaning send operation is performed by \s to \ob and forward\\\;\; is executed by \ob to \vp. The authorization functions evaluated to allow communication between \s and \vp include \\\;\;  $\mathrm{Auth_{send}}$(\s : S, \ob : TC) and $\mathrm{Auth_{forward}}$(\ob : TC, \vp : \vt). \end{tabular}}\\
\multicolumn{6}{l}{\begin{tabular}[c]{@{}l@{}}-- A source \s $\in$ S is allowed to communicate a message to vehicle \vp $\in$ \vt~if both the source and vehicle are associated with the\\\;\; same trusted cloudlet \ob $\in$ TC, and all the required policies are evaluated to make communication authorization decision. \\\;\; Formally, $\exists$ \ob $\in$ TC. ($\mathrm{Auth_{send}}$(\s : S, \ob : TC) $\land$ $\mathrm{Auth_{forward}}$(\ob : TC, \vp : \vt)) $\land$ (System-Wide Policies) = True \end{tabular}}
\\
\bottomrule
\end{tabularx}}
\end{table*}

%% file: Sections/Implementation.tex
\section{Implementation in AWS}\label{sec_imp}
In this section we present a proof of concept implementation of AB-ITS model in Amazon Web Services (AWS) \cite{aws}.  We use AWS IoT service along with AWS Greengrass \cite{grass} (to provide edge functionality) to setup a realistic environment where vehicles are simulated as AWS IoT things. In particular, these stand alone services are implemented as a Lambda function \cite{lambda} using Boto \cite{boto} which is AWS SDK for Python. It should be noted that in this implementation no long term GPS data coordinates of vehicles are collected in cloudlets. This reduces privacy concerns of end users and encourages adoption of the proposed model.

\subsection{Use Cases Overview}
US-DOT has proposed an extensive list of ITS applications \cite{its-app} which we have used to create our real life connect use-cases. Our implementation addresses trust, security and privacy issues concerning end users which must be satisfied before bringing ITS technology in practice. As most applications are safety related, we have considered accident and ice-on-road (tire slip) alerts as our running use-case along with real-time detection and prevention of rogue (or malicious) vehicles on road. In the use-cases, we have also shown how different entities (S, TC, \vt~etc.) fit in the formal model definitions.

\textbf{Accidental Safety and Ice-Threat :} Moving vehicles (S) can generate warnings for other vehicles (\vt) in their surrounding based on an event which they sensed or encountered. In our use-case, we consider `ice-threat' alerts based on a tire slip wherein vehicles are notified a warning, if any nearby vehicle `feels' it and broadcasts, after satisfying security policies implemented at the edge infrastructures (TC). These policies take into account: who is the source of alert, location of vehicle (ATT) and how many other vehicles encountered similar event, before forwarding (OP) these alerts to other nearby approaching vehicles (\vt). It is possible that a single vehicle (S) sends an ice-threat alert to associated cloudlet (TC), while other vehicles in the area sense no such movement.  Therefore the edge will be able to filter such malfunctioning or deliberate malicious attempt from the vehicle and also notify law enforcement and put that vehicle in  rogue vehicles list. Further, in case of an accident, alert messages will be generated and sent only to police or medical vehicles in the area. Based on the type of alerts and who generates it, policies are defined in the system to ensure trusted, anonymized and relevant notifications.
\input{Figures/comp}
\input{Tables/table_conf1}

\textbf{Compromised Rogue Vehicles :} Rogue vehicle either intentionally or due to sensor failure can send fake messages to other vehicles. Misbehaving and compromised vehicles must be detected in smart transportation and alerts must be issued immediately to discard the information sent by them. In our use-case, central cloud authority (S) informs edge infrastructures (TC) with a list of detected rogue vehicles and when any message is received by an edge from these vehicles, it is not forwarded to other vehicles. Further, law enforcement is informed about the location (ATT) of a rogue vehicle to prevent fake message dissemination. This approach prevents the need to update and publish revocation list to all vehicles eliminating the bandwidth and connectivity issues.

\input{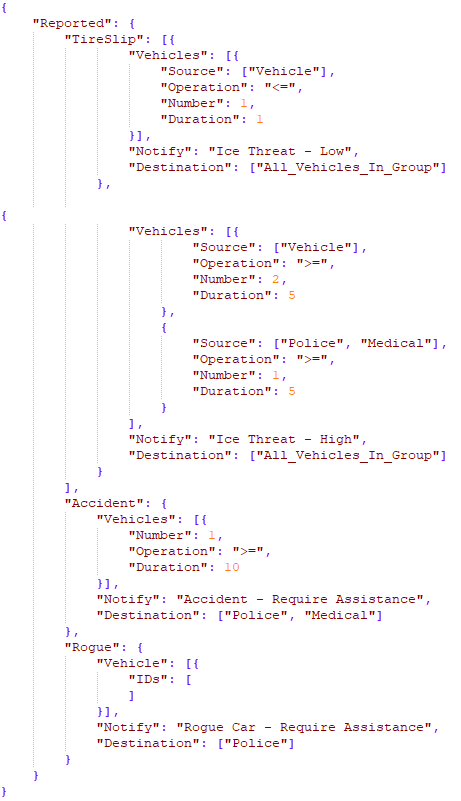}

\subsection{Proof of Concept}
We will first go over the system configuration along with implemented security policies defined in the cloudlet before we delve into more details of our developed prototype.

\textbf{System Architecture :} Figure \ref{fig-comp} represents system architecture along with different components implemented for our prototype. All the vehicles and static smart entities including edge infrastructures must be registered with a central cloud controller to ensure trusted authorized participating entities. Further, the controller also helps in the administrative phase (discussed later) which includes providing a list of edge infrastructures on designated path of moving vehicle. Once the registration is done and vehicles are sent a list of edge infrastructures, the vehicles publish and subscribe to secure (and reserved) MQTT topics created in each of cloudlets which get dynamically assigned based on vehicle current GPS coordinates.
It is also possible that the moving vehicle keeps on sending coordinates to the cloud and the controller lets them know the IP address of the nearby edge infrastructures to which the vehicle has to associate.
These cloudlets (represented as AWS Greengrass) hold the implemented security policies, a lambda function (similar to policy decision point - PDP \cite{hu2013guide}) for policy evaluation and the policy enforcement point (PEP) to check messages received, anonymize and filter them and based on the type of alert send them to relevant entities. It should be noted that only alert messages go through the enforcement point, whereas no alerts messages are discarded after logging. Table \ref{tab-2} lists different AWS system parameters to provide a better understanding of performance metrics shown later in this section.

\textbf{Security Policies :} We defined attributes based policies which are enforced at the edge, to check who is allowed to send messages, conditions when the message is forwarded to other vehicles and who are authorized recipients for different types of alerts in the system.
\input{Figures/gps}
\input{Figures/seq}
Various attributes can be included in policy but for the sake of simplicity we used only vehicle type to determine the source and destination of messages. As shown in Figure \ref{fig-policy}, security policies are listed in JSON format, where three types of alerts are being generated, `TireSlip', `Accident' and `Rogue' vehicle updates, as denoted by red rectangular boxes.
We defined separate set of conditions for each alert type. For example, in `TireSlip' alerts, it is first checked if it is generated (`Source' attribute) by a regular vehicle (specified by attribute value `Vehicle') or by law enforcement (`Police' or `Medical'). Policy then checks number of vehicles which created similar alerts (specified by "Number" attribute). Notification to other vehicles depends on how many alerts were generated or who is the source of alert. If the number of alerts are greater than or equal to 2 from regular vehicles, or even a single alert from police or medical vehicle, "Ice-threat High" notifications are sent to other associated vehicles of the cloudlet. However, if an alert is generated by one regular vehicle, "Ice Threat - Low" is sent for all member vehicles.
It must be noted that the sender vehicles and the receiving vehicle must be associated with the same cloudlet to exchange notifications, which also ensure relevance of alerts being received. Similarly, for accident use case, notification is only sent to nearby police vehicles and medical with assistance message. Here the source is not defined, since any smart entity including vehicle, or nearby smart road side sensor or a pedestrian can send message to police or medical vehicles. It is also possible that information about the vehicle including color, license plate number or other identifying information can be sent to law enforcement.
 Another important use case is to enable a central law enforcement that can regularly publish and update the list of rogue vehicles. This list for example, could help locate vehicles that have been stolen or implicated in amber alerts
 In the last part of our policy for `Rogue', vehicle IDs \texttt{Car-X, Car-Y, Vehicle-Z} are stated as rogue and any message from these vehicles is not forwarded. This is a dynamic policy as the list is periodically updated by a central authority. Also to extend the use-case, it is possible when an edge receives a message from a rogue vehicle, it can forward that information to nearby police along with vehicle information like license number and color. The defined policies are only for alert messages, and other `no alerts' messages are just checked by the policy and are logged and dropped without forwarding to any vehicle. Note that policies can also be implemented inside the smart vehicle as well to provide user privacy preference aware notifications, but are not implemented in our prototype.

\textbf{Implementation Details :} The implementation of our proposed solution involves two steps: the administrative phase and the operational phase. Administrative phase includes setup of cloudlets by city administration, setting up the boundaries for each cloudlet, dynamic assignment of moving vehicles to edge infrastructures, and attributes and alerts inheritance from edges to the member vehicles. To be part of ITS, vehicles and smart infrastructures need to have one time registration with central cloud which ensures that smart entities are trusted and benign.
Once registered, the moving vehicles can be provided with a mapped list of edges which will arrive in their designated route to which they are allowed to connect. As the vehicles get dynamically associated to different cloudlets, they are able to publish and subscribe to the reserved topics on each edge infrastructures. The operational phase consists of how these attributes and assignment to cloudlets ensure the relevance of alerts to the vehicles and how the edge deployed security policies are used to mitigate security and privacy concerns of users who are using AB-ITS system.

In our prototype, we demarcated a big geographic location area into several smaller regions and each region has a trusted cloudlet (TC) which serves all the smart entities in the region as shown in Figure \ref{fig gps}. We used a python script to simulate the movement of vehicles in the system, shown as green dots, which sends MQTT messages containing GPS coordinates to a central cloud. Service in cloud determines which edge cloudlets are in the surrounding area of the vehicle and then assigns the vehicle to the nearby cloudlets. Following is the sample MQTT payload sent by a moving vehicle to its shadow reserved topic \texttt{\$aws/things/`Vehicle-Name'/shadow/update} in the cloud for dynamic cloudlet assignment:

 \begin{itemize}[leftmargin=*]
  \item[] \texttt{\{"state": \{"reported": }
      \item[] \texttt{\;\;\;\;\;\;\;\;\;\;\;\;\;\;\;\;\;\;\;\;\;\;\;\;\;\;\; \{"Latitude": "28.1452683",}
  \item[] \texttt{ \;\;\;\;\;\;\;\;\;\;\;\;\;\;\;\;\;\;\;\;\;\;\;\;\;\;\; "Longitude":"-97.567259"\}\}\}}
\end{itemize}
As the path of vehicle is mostly known, these edge assignments can be pro-active in nature as well, mitigating the concern of cloud latency. In such a case, the cloud controller can send a list of edge infrastructures which will be on the designated path of the vehicle to get them associated when vehicles come in their range. It is also possible that these cloudlets have a wireless range and the vehicles which are in the range get automatically assigned to these cloudlets. A vehicle can associate to multiple cloudlets at a time based on their overlapping location. In Figure \ref{fig gps}, static smart objects like stop warning signs, road work ahead or other  infrastructures have fixed allocation to cloudlets, and the dotted lines represent predicted future cloudlets of vehicle along with current cloudlets by solid pink lines.

Once vehicles get assigned to nearby cloudlets, operational phase starts where the vehicles send messages to its shadow reserved topic (which gets created when the vehicle becomes member of a cloudlet) in their associated edges, which enforce security policies to ensure trusted and authorized alerts to nearby vehicles in near real time manner. In all the policies defined, privacy of the sender is well preserved as the messages do not contain any personal identifiable information and are anonymous. Following is a sample MQTT message sent by vehicle:
 \begin{itemize}[leftmargin=*]
  \item[] \texttt{\{"state":\{"reported": }
  \item[] \texttt{\;\; \{"Longitude": "29.472741982", }
  \item[] \texttt{\;\;\;\; "Latitude": "-98.50038363", }
  \item[] \texttt{\;\;\;\; "Time": "2019-03-19 11:27:40.237734", }
  \item[] \texttt{\;\;\;\; "Velocity": "30", "Direction": "north", }
  \item[] \texttt{\;\;\;\;\;\;"Elevation": "650", "Posit. Accuracy":}
  \item[] \texttt{\;\;\;\;  "5", "Steering Wheel Angle": "0",}
  \item[] \texttt{\;\;\;\; "Alert": myAlert\}\}\}}
\end{itemize}
In this message, beside BSM \cite{bsm} attributes, an attribute "Alert" also exists, which defines what kind of alert has been sent from the vehicle to cloudlets. For our use-cases, it can be an "Accident", "Tireslip", or "Null" value where Null signifies no alert. Once the message is received by cloudlet, and is checked against the policies, the edge infrastructure forwards the following Tireslip alert message to a generic topic \texttt{test/devices} to which the vehicles subscribe when they become member of the edge.
 \begin{itemize}[leftmargin=*]
\item[] \texttt{\{"message": "Ice Threat - Low', }
\item[] \texttt{ 'myEvent': '2019-03-19 10:56:15.921834'"\}}
\end{itemize}
In case of accident alert following message:
\begin{itemize}[leftmargin=*]
\item[] \texttt{\{"message":"Accident- Require Assistance',}
\item[] \texttt{ 'myEvent': '2019-03-19 11:27:40.237734'"\}}
\end{itemize}
is sent to topic \texttt{test/medical} and \texttt{test/police} to which nearby medical and police vehicles are subscribed respectively. Note that event time has also been added to messages, to ensure when the message  is not obsolete. Similarly, for updating the rogue vehicle list from the transportation authority via central cloud to the edge infrastructures, message
\begin{itemize}[leftmargin=*]
\item[] \texttt{\{"Alert": myAlert, "myVehicle": myVehicle\}}
\end{itemize}
is sent to \texttt{test/Rogue-Vehicle} topic. In this message, 'myAlert' variable can be \texttt{ADD}, \texttt{DELETE} or \texttt{LIST} operation, and 'myVehicle' can hold the vehicles to be added or deleted. In case of list operation, 'myVehicle' attribute value is NULL.
The complete sequence of events for the administrative and operational phase in cloudlet supported ITS is shown in Figure \ref{fig_seq}.
\input{Figures/perform}
\subsection{Performance Metrics and Discussion}

We evaluated the performance of our proposed AB-ITS model in AWS and provide metrics for the use-cases in proof of concept. We first calculate the execution time for the proposed policy enforcer to evaluate the attribute based security polices (shown in Figure \ref{fig-policy}) against the number of vehicles associated with a cloudlet and scaling the number of messages sent per vehicle per second. In Figure \ref{fig-perform} (a) and (b), as the number of vehicles increase (along x axis) with more messages being sent, the enforcer takes more time to evaluate the polices and impact performance. This enforced policy engine in cloudlet has the worse case execution time less than 200 microseconds, for any number of messages sent per second (from 1 to 20) by vehicles which could range from 1 to 50. In case of no-alerts, this execution time will be zero as the policies will not be evaluated.
Total trip time performance of our model includes time at which vehicle generates an alert till it is received by target vehicles which includes the policy evaluation time. As shown in Table \ref{tab-acc} and \ref{tab-tire}, the total trip time is within the permissible limits ($\sim$100 ms \cite{Xu:2004:VSM:1023875.1023879}) for most of the case scenarios. However, the trip time goes beyond the limits when 50 vehicles get associated to single edge cloudlet at one time. The variation in total trip time is due to network traffic and latency, but the average and standard deviation infer that the performance is very comparable to peer to peer ITS. It should be noted that the extra overhead induced by policy execution (in microseconds) is very negligible as compared to the total trip time (in milliseconds). In our approach MQTT protocol has been used, therefore, if some one does not want use DSRC due to cost of transmitter and receiver, our approach can still work with the traditional IoT MQTT based communication based on LTE, 5G or WiFi connectivity.
\input{Tables/perform}
We understand that there may be hundreds of vehicles during heavy traffic time, therefore, to scale the system and accommodate all vehicles we can install more cloudlets and infrastructure devices in busy areas that will reduce the number of vehicles which will get associated with single cloudlet at a time. This implementation in AWS showcases the practical viability and use of fine grained polices in context of intelligent transportation system, without the need to capture data points from real world traffic. It must be also noted that, AWS Greengrass has limit of 200 devices per Greengrasss group, which means maximum number of vehicles which can be associated can not be more than 200. We can add more cloudlets in the system which can cater to higher population of vehicles and smart entities.
As mentioned earlier, this proposed cloudlet supported V2V and V2I complements the current DSRC approach and is not considered a replacement.







%% file: Figures/comp.tex
\begin{figure}[t]
\centering
\includegraphics[scale = .65]{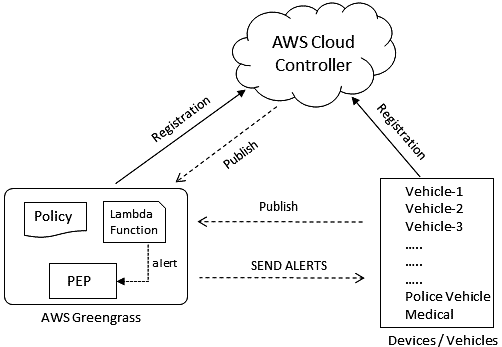}
\caption{System Architecture}
\label{fig-comp}
\end{figure} 

%% file: Tables/table_conf1.tex
\begin{table}[]
\caption{AWS Setup Parameter Information}
\label{tab-2}
\setlength{\tabcolsep}{5pt} 
\renewcommand{\arraystretch}{1.1} 
\begin{tabular}{cc}
\hline
\multicolumn{1}{|c|}{Message Queue Size}                                                                              & \multicolumn{1}{c|}{2.5 MB}                                                                     \\ \hline
\multicolumn{1}{|c|}{Number of AWS Greengrass Groups}                                                                 & \multicolumn{1}{c|}{4}                                                                          \\ \hline
\multicolumn{1}{|c|}{Range of Vehicles per groups}                                                                    & \multicolumn{1}{c|}{1 -- 50}                                                                    \\ \hline
\multicolumn{1}{|c|}{\begin{tabular}[c]{@{}c@{}}Published Range of Messages \\ (per second per vehicle)\end{tabular}} & \multicolumn{1}{c|}{1 -- 20}                                                                    \\ \hline
\multicolumn{1}{|c|}{Greengrass Server Configuration}                                                                 & \multicolumn{1}{c|}{8 VCPUs, 4 GB RAM}                                                          \\ \hline
\multicolumn{1}{|c|}{Simulated Vehicles Server Config.}                                                                & \multicolumn{1}{c|}{2 VCPUs, 4 GB RAM}                                                          \\ \hline
\multicolumn{1}{|c|}{Average Network traffic (50 vehicles)}                                                                 & \multicolumn{1}{c|}{ 255 Kbps}                                                          \\ \hline
\multicolumn{1}{|c|}{Network Capacity of Interface}                                                                & \multicolumn{1}{c|}{1.84 Mbps}                                                          \\ \hline
\multicolumn{1}{l}{}                                                                                                  & \multicolumn{1}{l}{}
\end{tabular}
\end{table}

%% file: Figures/Policy.tex
\begin{figure}[t]
\centering
\includegraphics[scale = .70] {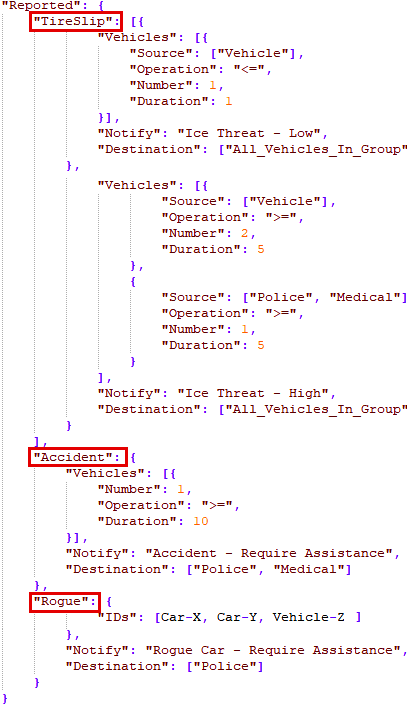}
\caption{Implemented Security Policies}
\label{fig-policy}
\end{figure} 

%% file: Figures/gps.tex
\begin{figure}[t]
\centering
\includegraphics[scale=.42]{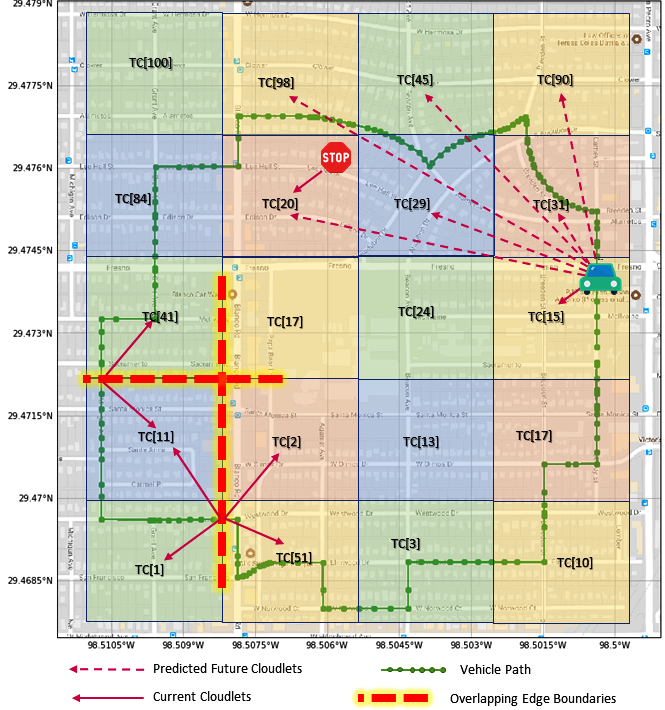}
\caption{Moving Vehicle Cloudlets Association}
\label{fig gps}
\end{figure} 

%% file: Figures/seq.tex
\begin{figure*}[t]
\centering
\includegraphics[scale = .60]{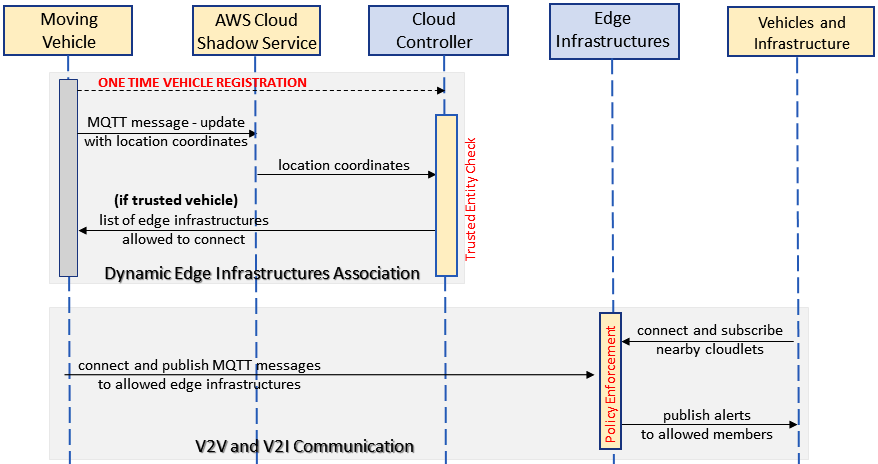}
\caption{Sequence Diagram for Cloudlets Supported V2V and V2I Communication}
\label{fig_seq}
\end{figure*} 

%% file: Figures/perform.tex
\begin{figure}[!tbp]
\centering
\subfloat[Accident Use Case]{\includegraphics[scale = .37]{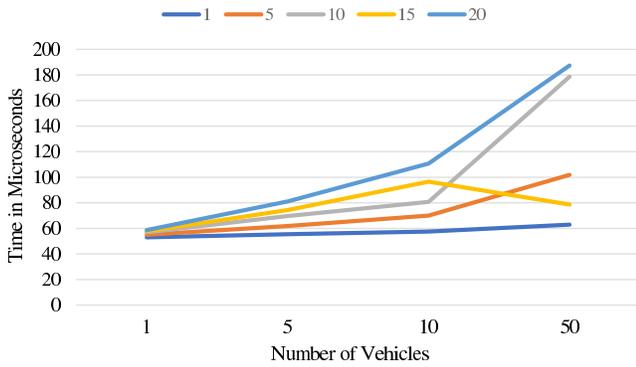}}
\label{fig-acc}
\hfill
\subfloat[Tire-Slip Use Case]{\includegraphics[scale = .37]{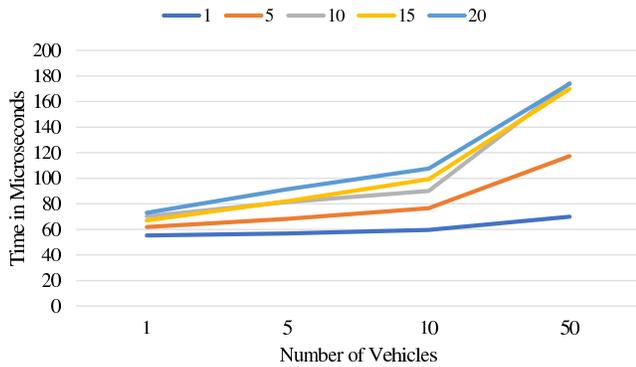}}
\label{fig-tire}
\caption{Policy Evaluation Time}
\label{fig-perform}
\end{figure} 

%% file: Tables/perform.tex
\begin{table}[]
\caption{Total Trip Time for Accident}
\label{tab-acc}
\begin{tabular}{|c|c|c|c|c|}
\hline
{Msg. per Sec, Vehicles}
 & \textbf{1} & \textbf{5} & \textbf{10} & \textbf{50} \\ \hline
\textbf{1} & 71.72 & 23.53 & 32.45 & 39.85 \\ \hline
\textbf{5} & 18.94 & 79.69 & 78.87 & 69.11 \\ \hline
\textbf{10} & 30.73 & 73.73 & 28.57 & 83.89 \\ \hline
\textbf{15} & 18.01 & 22.31 & 30.06 & $\sim$ \\ \hline
\textbf{20} & 18.04 & 34.40 & 65.82 & $\sim$ \\ \hline
\textbf{Average} & 31.49 & 46.73 & 47.15 & 64.28 \\ \hline
\textbf{Standard Deviation} & 23.13 & 27.85 & 23.49 & 22.42 \\ \hline
\end{tabular}
\end{table}

\begin{table}[]
\caption{Total Trip Time for TireSlip}
\label{tab-tire}
\begin{tabular}{|c|c|c|c|c|}
\hline
{Msg. per Sec, }{Vehicles}
 & \textbf{1} & \textbf{5} & \textbf{10} & \textbf{50} \\ \hline
\textbf{1} & 47.44 & 56.24 & 89.78 & 55.72 \\ \hline
\textbf{5} & 104.23 & 99.27 & 56.76 & 85.26 \\ \hline
\textbf{10} & 43.38 & 44.07 & 51.49 & $\sim$ \\ \hline
\textbf{15} & 66.43 & 44.04 & 51.32 & $\sim$ \\ \hline
\textbf{20} & 42.76 & 45.74 & 85.40 & $\sim$ \\ \hline
\textbf{Average} & 60.85 & 57.87 & 66.95 & 70.49 \\ \hline
\textbf{Standard Deviation} & 26.10 & 23.69 & 19.03 & 20.89 \\ \hline
\end{tabular}
\end{table}

%% file: Sections/Summary.tex
\section{Summary}\label{sec_summary}

This research work proposes a cloudlet assisted secure V2V and V2I communication in intelligent transportation system, which ensures trusted and reliable messages exchange among moving entities on road. We introduce the novel notion of dynamic edge associations in which the smart entities get connected to different pre-installed cloudlets on road, which help them relay the basic safety messages and perform the needed filtering and reduces privacy concerns of the users. These cloudlets can anonymize the messages, ensure trustworthiness and ensure their relevance to entities which receive them. We also present the formal model which specifies attributes based polices for V2V and V2I communication. Several use-cases of ITS have been discussed along with implementation in Amazon Web Services (AWS). Performance has been evaluated against time taken to evaluate the polices in cloudlets and the total trip time from the moment message is generated till it gets received and relayed by the cloudlets. In future work we would incorporate additional privacy preserving approaches wherein the exact location GPS coordinates are not required to be shared with cloud. The work can be complemented using homomorphic encryption or other similar approaches which will further mitigate privacy concerns of the users.